\documentstyle[12pt]{article}
\newcommand{\be}{\begin{equation}}
\newcommand{\ee}{\end{equation}}

\newcommand{\bear}{\begin{eqnarray}}
\newcommand{\eear}{\end{eqnarray}}
 
\newcommand{\parl}{\left(\hspace{-0.4cm}\begin{array}{c}\\ 
\end{array}\right.} 
\newcommand{\parr}{\left)\hspace{-0.4cm}\begin{array}{c}\\ 
\end{array}\right.} 
\newcommand{\corl}{\left[\hspace{-0.4cm}\begin{array}{c}\\ 
\end{array}\right.} 
\newcommand{\corr}{\left]\hspace{-0.4cm}\begin{array}{c}\\ 
\end{array}\right.} 
\newcommand{\corrl}{\left[\hspace{-0.4cm}\begin{array}{c}\\
\\ 
\end{array}\right.} 
\newcommand{\corrr}{\left]\hspace{-0.4cm}\begin{array}{c}\\ 
\\
\end{array}\right.} 
\newcommand{\llavl}{\left\{\hspace{-0.4cm}\begin{array}{c}\\
\\ \end{array}\right.} 
\newcommand{\llavr}{\left\}\hspace{-0.4cm}\begin{array}{c}\\ 
\\ \end{array}\right.}

\def\vi{\langle h_i^{0r}\rangle}
\def\vj{\langle h_j^{0r}\rangle}
\def\IJMPA #1 #2 #3 {Int.~J.~Mod.~Phys.~{\bf A#1}\ (19#2) #3}
\def\MPLA #1 #2 #3 {Mod.~Phys.~Lett.~{\bf A#1}\ (19#2) #3}
\def\NPB #1 #2 #3 {Nucl.~Phys.~{\bf B#1}\ (19#2) #3}
\def\PLB #1 #2 #3 {Phys.~Lett.~{\bf B#1}\ (19#2) #3}
\def\PR #1 #2 #3 {Phys.~Rep.~{\bf#1}\ (19#2) #3}
\def\PRD #1 #2 #3 {Phys.~Rev.~{\bf D#1}\ (19#2) #3}
\def\PTP #1 #2 #3 {Prog.~Theor.~Phys.~{\bf #1}\ (19#2) #3}
\def\PRL #1 #2 #3 {Phys.~Rev.~Lett.~{\bf#1}\ (19#2) #3}
\def\RMP #1 #2 #3 {Rev.~Mod.~Phys.~{\bf#1}\ (19#2) #3}
\def\ZPC #1 #2 #3 {Z.~Phys.~{\bf C#1}\ (19#2) #3}

\begin{document}

\begin{titlepage}

\title{\bf New limits on the mass of neutral Higgses in General Models}

\author{
{\bf D. Comelli}\thanks{Work supported by Ministerio de Educaci\'on y 
Ciencia (Spain).}\\ Instituto de F\'{\i}sica Corpuscular - IFIC/CSIC \\
Dept. de F\'{\i}sica Te\`orica, Universidad de Val\`encia\\
46100 Burjassot, Val\`encia. Spain\\
and\\
{\bf J.R. Espinosa} \thanks{Work supported by the Alexander-von-Humboldt 
Stiftung.} \\ Deutsches Elektronen Synchrotron DESY. \\
Notkestrasse 85.\ \ 22603 Hamburg. Germany}

\date{} 
\maketitle
\vspace{.5cm}
\def\baselinestretch{1.15}
\begin{abstract}
In general electroweak models with weakly coupled (and 
otherwise arbitrary) Higgs sector there 
always exists in the spectrum a scalar state with mass controlled by the 
electroweak scale. A new and simple recipe to compute an analytical 
tree-level upper bound on the mass of this light scalar is given. 
We compare this new bound with similar ones existing in the literature 
and show how to extract extra information on heavier neutral scalars 
in the spectrum from the interplay of independent bounds. Production of 
these states at future colliders is addressed and the implications for 
the decoupling  limit in which only one Higgs is expected to remain light 
are discussed. 
\end{abstract}
\vspace{2cm}
\leftline{July 1996}

\thispagestyle{empty}

\vskip-20.5cm
\rightline{{\bf DESY 96-133}}
\rightline{{\bf FTUV/96-43}}
\rightline{{\bf IFIC/96-51}}
\rightline{{\bf IEM--FT--136/96}}
\rightline{{\bf hep-ph/9607400}}
\vskip3in

\end{titlepage}

\def\baselinestretch{1.1}

In the Standard Model the mass of the Higgs boson is an unknown parameter
waiting to be measured and all the information we can provide about it to 
guide its search is precious. Lower bounds on it can be obtained requiring
the electroweak minimum of the Higgs potential to be stable or 
metastable and upper bounds can also be found imposing that the Standard 
Model remains perturbative up to some high energy scale $\Lambda$. This 
last point is made possible by the fact that
the squared mass $m_h^2$ of the Higgs boson is of the form $\lambda v^2$
with $v$ fixed by the gauge boson masses and $\lambda$ being the (not 
asymptotically free) scalar
self coupling. Imposing that $\lambda$ remains perturbative below 
$\Lambda$ one gets an upper bound on $m_h$, especifically $m_h<180-200\ GeV$ 
for $\Lambda=10^{16}\ GeV$. 

In more general models of electroweak breaking one typically has more 
physical Higgs bosons (scalars, pseudoscalars, charged,...). The Higgs
potential will contain more mass parameters and the mass spectrum of 
the Higgs sector will be more complicated. However, in any model of 
electroweak breaking with weakly coupled Higgs sector at least one of 
the physical Higgs bosons has a mass controlled by the electroweak scale 
\cite{lanwel,wel}, i.e. there is an upper bound on the squared
mass of the lightest Higgs boson of the form $\lambda v^2$, with $\lambda$
some combination of quartic Higgs self couplings and $v$ of the order
of the electroweak scale. This means that in the most general model one
can always find an upper bound on (at least) one Higgs scalar imposing 
that the theory remains perturbative up to some high energy scale.

In section 1 we present a novel proof of this fact deriving a very simple
mass bound, different and independent of those presented in 
\cite{lanwel,wel}. This bound can be applied in many contexts and models. 
As a relevant example we derive in section 2 the form of this bound in 
general supersymmetric Standard models. The new bound is compared with 
the old ones of refs.~\cite{lanwel,wel} in section~3 while section~4 is 
devoted to extract some implications of the interplay between different 
bounds. Finally, section~5 examines the production cross section of these 
light neutral scalars in general models.
\newline

{\bf 1. Upper limit on the mass of the lightest neutral scalar}
\vspace{0.5cm}

The proof goes as follows: let $\Phi_j$ be the (generically complex) 
neutral scalar fields in the model (i.e. all those fields susceptible of 
taking a VEV). They belong to $(2T_j+1)-SU(2)_L$ multiplets and have 
hypercharges $Y_j$. We write:
\be
\Phi_j=\frac{1}{\sqrt{2}}\corl \phi_j^{0r} + i \phi_j^{0i}\corr .
\ee
After the breaking of $SU(2)_L\times U(1)_Y$ down to $U(1)_{em}$ we 
define the real field $\phi^0$ as:
\be
\label{phi0}
\phi^0=\frac{1}{\langle\phi^0\rangle}\sum_j 
{}^{'}\corl\langle\phi_j^{0r}\rangle
\phi_j^{0r}+\langle\phi_j^{0i}\rangle\phi_j^{0i}\corr,
\ee
with
\be
\langle\phi^0\rangle^2=2\sum_j {}^{'}|\langle\Phi_j\rangle|^2,
\ee
and the primed sum extends only to $2T_j$ odd fields (i.e. doublets, 
4-plets, etc. but not singlets, triplets,...). The reason for this will 
be clear later on. Definition (\ref{phi0}) ensures that $\phi^0$ will be 
the only $2T_j$ odd field having a non-zero VEV. All other $2T_j$ odd 
fields orthogonal to $\phi^0$ have zero VEV by construction.

The structure of the electroweak breaking is represented pictorially in 
Fig.~1. The breaking can be decomposed in $2T_j$ even and odd 
parts. The $2T_j=0$ VEV (coming from singlets) has been drawn separately 
because it does not break $SU(2)\times U(1)$.
We can calculate the tensor $M^2_{ij}$ of (scalar neutral) mass 
excitations in the 
vacuum $\langle\phi^0\rangle$ and in particular we will be interested in 
excitations along the $\phi^0$-direction: call the corresponding squared 
mass $M^2_{\phi^0}\equiv\langle\phi^0 | M^2 | \phi^0\rangle$. This mass 
provides an upper bound on the mass of the lightest neutral scalar of the 
theory:
if $M^2_{\phi^0}$ is a true eigenvalue of $M^2$ then the bound is 
saturated. If it is not a true eigenvalue then $\langle\phi^0 | M^2 |
 \phi^0\rangle$ is not stable under small perturbations in the field 
direction $\phi^0$, that is, we can find some field direction along which 
this mass is reduced and so, some eigenvalue of $M^2$ lies below 
$M_{\phi^0}^2$. (And that eigenvalue is not the neutral Goldstone: note 
that $M^2_{\phi^0}$ will be stable under perturbations along the 
"Goldstone direction").
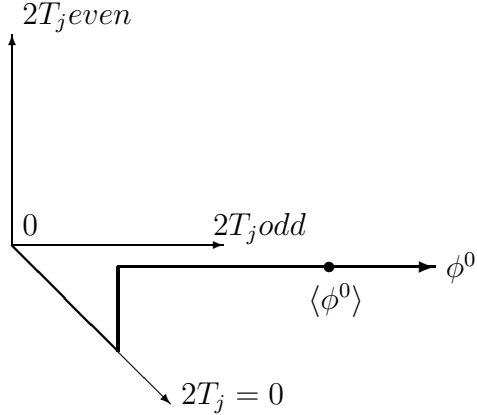
\begin{figure}
\begin{center}
\setlength{\unitlength}{0.4pt}
\begin{picture}(500,500)(0,450)
\put(60,640){\makebox(0,0)[lb]{$0$}}
\put(60,840){\makebox(0,0)[lb]{$2T_j even$}}
\put(240,640){\makebox(0,0)[lb]{$2T_j odd$}}
\put(210,480){\makebox(0,0)[lb]{$2T_j=0$}}
\put(330,570){\makebox(0,0)[lb]{$\langle \phi^0 \rangle$}}
\put(460,600){\makebox(0,0)[lb]{$\phi^0$}}
\thinlines
\put(50,630){\vector(0,1){200}}
\put(50,630){\vector(1,0){200}}
\put(50,630){\vector(1,-1){150}}
\thicklines
\put(350,610){\circle*{10}}
\put(50,630){\line(1,-1){100}}
\put(150,530){\line(0,1){80}}
\put(150,610){\vector(1,0){300}}
\end{picture}
\end{center}
\caption{\footnotesize The structure of electroweak breaking}
\end{figure}

Having proved that the quantity $M^2_{\phi^0}$ is a bound on the mass of 
the lightest neutral scalar in the theory we now calculate its general form. 
For this we need to consider the scalar potential along the direction 
$\phi^0$. Towards this end, consider the (multiplicative) discrete 
symmetry 
\be
\label{sym}
\Phi_j\rightarrow (-1)^{2T_j}\Phi_j,
\ee
(that can be extended to all the fields in the theory). The Lagrangian 
being a $SU(2)_L$-singlet is invariant under (\ref{sym}). When the 
$2T_j$-even fields take a VEV, the symmetry remains unbroken. So, 
$V(\phi^0)$ must respect also this symmetry. As $\phi^0\rightarrow -\phi^0$ 
under the transformation (\ref{sym}), the form of $V(\phi^0)$ is 
restricted to be, at tree level:
\be
\label{pot}
V(\phi^0)=V_0 - \frac{1}{2}m^2\phi^0\phi^0 
+\frac{1}{8}\lambda(\phi^0\phi^0)^2,
\ee
with cubic and linear terms forbidden. To see 
how this comes out in detail consider a linear term $\delta 
V=\kappa\phi^0$ in the potential. If this comes from a full scalar 
potential $SU(2)_L$ invariant, then $\kappa$ should be the VEV of a field 
or combination of fields transforming with $2T$ odd. The first 
case ($\kappa\sim\langle\phi_j\rangle$) is not possible because 
all $2T_i$ odd fields orthogonal to $\phi^0$ have zero VEV. The second 
case ($\kappa\sim\langle\phi_i\phi_j...\rangle$) is also 
impossible because some of the fields $\phi_i$ appearing in $\kappa$ must 
transform as $2T_i$ odd if $\kappa$ itself is $2T$ odd (the combination
of $2T_i$ even fields produces only $2T$ even fields), and again those 
have been already projected into $\phi^0$. A similar argument applies to 
the cubic terms in $V(\phi^0)$.

Then, from eq. (\ref{pot}), using the condition $\partial 
V/\partial\phi^0=0$ at the minimum $\langle\phi^0\rangle=v$ we get
\be
\label{bound}
M^2_{\phi^0}=\lambda v^2,
\ee
where $v$ is related to the electroweak scale through the gauge boson 
masses [it will be different from model to model but always $v^2\leq 
(\sqrt{2}G_F)^{-1}$], and $\lambda$ is some 
combination of quartic couplings of the theory so that it is sensible to 
require it to be perturbative.

We have proved our $SU(2)$-based theorem extracting from $SU(2)_L$ a 
discrete 
symmetry to restrict the form of the potential in some field direction. 
We can follow the same procedure for any spontaneously broken $U(1)$ 
(global or local). Our $U(1)$ derivation differs from a similar one 
given in \cite{bucce}, the conclusion being the same. Let us concentrate 
in the neutral fields $\phi_i$ which have a non-zero $U(1)$ charge $Q_i$ 
(the $Q_i$'s are some fractional numbers, $Q_i=n_i/d_i$) and take a VEV. 
We can always rescale the charges multiplying them by L, the least 
common multiple of the denominators $d_i$ in such a way that the new 
charges $Q_i'=LQ_i$ are integers. The potential for the fields 
$\phi_i$ will be invariant under the discrete symmetry
\be
\label{symu1}
\phi_i\rightarrow (-1)^{Q_i'}\phi_i.
\ee
We can decompose the VEVs in a singlet, an odd and an even part 
according to its properties under (\ref{symu1}). Then, a mass bound can be 
derived looking in the field direction determined by
\be
\phi^0=\frac{1}{\langle\phi^0\rangle}\sum_j 
{}^{'}\corl\langle\phi_j^{0r}\rangle
\phi_j^{0r}+\langle\phi_j^{0i}\rangle\phi_j^{0i}\corr,
\ee
where the primed sum extends to odd $Q_i'$ fields only. 

We can apply this result to the breaking of $U(1)_Y$. As we are 
always assuming that only neutral fields take a VEV we have the relation 
$Y_i=-T_{3i}$ for the i-th field. The $T_{3i}$ are integers or half 
integers. If some half-integer $Y_i$ field is taking a VEV [e.g. some 
$SU(2)_L$ doublet, as required to give masses to quarks] then L=2 and the 
discrete symmetry (\ref{symu1}) is \be
\phi_i\rightarrow (-1)^{LY_i}\phi_i=(-1)^{2T_{3i}}\phi_i=(-1)^{2T_i}\phi_i,
\ee
and this is nothing but the old $SU(2)$-based discrete symmetry 
(\ref{sym}) so that no further information can be extracted from the 
$U(1)_Y$ breaking\footnote{In  case that only integer $Y_i$ fields develop a 
VEV, $L=1$ and now the mass bound can be derived from the $U(1)_Y$ breaking 
while the $SU(2)_L$ breaking gives no information.}.
\newline

{\bf 2. Supersymmetric models}
\vspace{0.5cm}

In the Minimal Supersymmetric Standard Model the Higgs sector contains 
two Higgs doublets ($H_1,\,H_2$) so that, after electroweak symmetry 
breaking ($\langle H_1^0\rangle=v_1/\sqrt{2},\,\langle H_2^0\rangle=
v_2/\sqrt{2}$) the spectrum of physical Higgses consists of two scalars, 
$h^0$ and $H^0$, one pseudoscalar, $A^0$, and a pair of charged Higgses 
$H^{\pm}$. On the other hand supersymmetry and gauge invariance restrict 
the interactions in the Higgs sector in such a way that the mass spectrum 
is quite constrained and at tree level it is completely determined by just 
two parameters: $\tan\beta=v_2/v_1=\langle H_2^0\rangle/\langle H_1^0
\rangle$ and the mass of one of the Higgses, conventionally taken to be 
$m_{A^0}$.

As is well known, in the Minimal Supersymmetric Standard Model the tree
level mass $m_h$ of the lightest scalar $h^0$ is bounded by 
$M_Z | \cos2\beta |$.
Radiative corrections to the mass of this Higgs boson can be large if
the mass of top and stops is large, and the tree level bound can be 
spoiled. But, even if the lightest Higgs can scape detection at LEP-II 
its mass is always of the order of the electroweak scale (the 
dependence on the soft breaking scale is only logarithmic). 

The situation in extended supersymmetric models is somewhat 
qualitatively different.  
An analytical upper bound on the tree-level mass of the lightest Higgs 
boson is known for very general Supersymmetric Standard Models with
extended Higgs or gauge sectors \cite{eq,kane}. This bound depends on the 
electroweak
scale (given by $M_Z$) and on the new Yukawa or gauge couplings that
appear in the theory, which are not fixed by experiment as in the MSSM. 
Numerical bounds can be obtained by putting
limits on these unknown parameters ({\it e.g.} assuming that the theory 
remains perturbative up to some high scale). 
These bounds are typically greater than the MSSM bound but still of 
order $M_Z$, for example \cite{eq} one gets $m_h<155\ GeV$ for 
$\Lambda=10^{16}\ GeV$. 

After the results of section 1 it should be clear that there is nothing 
special 
concerning supersymmetry in what respects to the bound on the lightest 
Higgs boson: the existence of the mass bound controlled by the 
electroweak scale follows directly from gauge invariance without the need 
to advocate supersymmetry [the role of supersymmetry is to fix in some 
cases ({\em e.g.} the MSSM) the quartic couplings appearing in the bound]. 

Following section~1, the derivation of the tree-level upper 
bound on the mass of the lightest Higgs boson is greatly simplified:
Consider the most general supersymmetric standard model assuming for 
simplicity that  $CP$ is conserved and all $\phi_j^{0i}=0$. Define the 
field $\Phi^0$ according to (\ref{phi0}) and calculate the quartic term 
$(\phi^0)^4$ in the potential. This comes from two contributions:

{\em i)} F-Terms: write the superpotential for the neutral fields in 
terms of $\Phi^0$ plus orthogonal fields. The only terms that will 
contribute to $(\phi^0)^4$ in the potential are of the form
\be
\label{superpot}
\delta f = \sum_{ijk} \lambda_{ijk}\varphi_i^0\Phi_j^0\Phi_k^0 =
\sum_{ijk} \lambda_{ijk}c_jc_k\varphi_i^0\Phi^0\Phi^0+...\equiv
\sum_i \lambda_i\varphi_i^0\Phi^0\Phi^0+...  , 
\ee
where now the fields $\varphi_i$ are $2T_i$ even and $c_j=\langle 
\Phi_j^0\rangle/\langle\Phi^0\rangle$ appear from the projection of 
odd-fields onto $\Phi^0$. Then it results
 \be \label{yuk}
\delta V=\frac{1}{4}\sum_i\lambda_i^2 (\phi^0\phi^0)^2.
\ee

{\em ii)} D-terms: The contribution from $SU(2)_L\times U(1)_Y$ is easily 
calculable and gives
\be
\label{dter}
\delta 
V=\frac{1}{8}(g^2+g'^2)\parl\sum_j{}^{'}Y_jc_j^2\parr^2(\phi^0\phi^0)^2,
\ee
where $c_j=\langle\phi_j^0\rangle/\langle\phi^0\rangle$.
The contribution from other gauge groups that might be present can also 
be added without difficulty following the same procedure.

Putting together the contributions coming from (\ref{yuk}) and 
(\ref{dter}) the bound on the mass of the lightest Higgs boson can be 
written as
\be
\label{genbou}
m_h^2\leq(g^2+g'^2)\parl\sum_j{}^{'}Y_jc_j^2\parr^2v_o^2 + 
2\sum_i\lambda_i^2v_o^2 ,
\ee
where $v_o=\langle\phi^0\rangle$. The subscript is meant to remind that 
only odd-fields contribute to $v_o$.

Previous studies of this kind in general supersymmetric models analyzed 
$2\times 2$ submatrices of the neutral scalar matrix. We have shown that 
this is an unnecessary complication and moreover we can see that the naive 
application of this technique can fail in some cases (see Appendix). In 
the light of the case presented in the Appendix one can wonder whether the 
universal upper bound calculated numerically for $M_h$ in models with an
arbitrary Higgs sector \cite{eq} remains valid or gets modified (after all 
the technique of $2\times 2$ submatrices was used to extract the analytic 
bound). The key point here is that, to affect the derivation of 
the analytical bound 
obtained in \cite{eq} one needs some even (non-singlet) fields with 
non-zero VEV. In such a case $v_o$ is proportionally reduced with a 
corresponding decrease in the bound on $M_h$. For that reason, when 
looking for the universal upper bound in models with arbitrary Higgs 
sector one has to concentrate only in those cases with $v_e=0$, as was 
done in \cite{eq,espinosa}.
\newline

{\bf 3. Comparison with existing bounds}
\vspace{0.5cm}

Bounds similar to the one we presented in Section 1 have been previously 
derived  in the literature \cite{lanwel,wel}. All of them have the form
$\lambda v^2$, ($\lambda$ being some quartic scalar coupling and $v$ some
VEV controlled by the measured gauge boson masses) but are generically 
different as they arise from looking into different field space 
directions. We review here those 
bounds in a unified manner that should help the comparison between them.

All the bounds are obtained from an inequality of the form
\be
\label{primbound}
\langle\varphi_a | M^2 | \phi_a \rangle \leq \lambda_a v_a^2,
\ee
where the index $a$ just runs over different bounds. The states 
$\varphi_a$ and $\phi_a$ are some real fields that can be normalized to 
satisfy $\langle \varphi_a | \phi_a \rangle = 1$ (and in most cases 
$\varphi_a=\phi_a$). The final form of the bounds follow from 
(\ref{primbound}) immediately\footnote{If $\phi_a\neq\varphi_a$ the 
condition $\langle\varphi_a|A\rangle\langle A|\phi_a\rangle\geq 0$ must 
be satisfied.}: 
\bear \label{intbound}
\lambda_a v_a^2 &\geq& \langle \varphi_a | M^2 | \phi_a \rangle = \sum_{A,B}
\langle \varphi_a | A \rangle M_A^2 \delta_{AB} \langle B | \phi_a\rangle 
\nonumber\\
&=&\sum_A M_A^2 \langle \varphi_a | A \rangle  \langle A | \phi_a\rangle \geq
M_h^2 \langle\varphi_a|\phi_a\rangle=M_h^2,
\eear
where the $|A\rangle$'s form a basis of neutral scalar mass eigenstates. The 
form 
(\ref{primbound}) will be useful for the discussion of the decoupling 
limit in Section~4. In particular, knowledge of the field directions 
$\varphi_a$ and $\phi_a$ will be needed. We give them below together 
with the different mass bounds on $M_h^2$. To simplify we will assume 
that CP is conserved in the Higgs sector and all the VEVs can taken to be 
real. 

{\bf Bound 1.} It was obtained in ref.~\cite{lanwel} looking at the 
potential in the direction 
\be
\label{field1}
\varphi_1\equiv\phi_1=\frac{1}{v_1}\sum_i\langle h_i^{0r}\rangle h_i^{0r},
\ee
with 
\be
v_1^2=\sum_i\langle h_i^{0r}\rangle^2=v_o^2+v_e^2.
\ee
Here the sum runs over all non-singlets, even or odd, as indicated by the 
last equality.
 
The tree-level potential $V(\phi_1)$ has no linear terms in $\phi_1$ but 
cubic ones are now allowed:
\be
V(\phi_1)=V_0 + m^2\phi_1^2 + \sigma\phi_1^3 +\frac{1}{8}\lambda_1\phi_1^4.
\ee
Nevertheless, a bound can be obtained assuming that the electroweak 
vacuum is the deepest one (the bound would not apply if the vacuum were 
metastable at tree level)  and takes the form
\be
\label{bound1}
M_h^2\leq\lambda_1(v_o^2+v_e^2).
\ee

{\bf Bounds 2,3,4.} These three bounds were obtained in ref.~\cite{wel} 
making use of gauge invariance to relate different order derivatives of the 
effective potential. In this way second derivatives (masses) can be 
connected to fourth derivatives (quartic couplings) and the bounds follow.
In fact it is remarkable that they apply to very general non-polynomial 
potentials although 
here we will restrict ourselves to tree-level polynomial potentials.

The field directions in (\ref{primbound}) are
\bear
\label{fields234}
\varphi_2\equiv\phi_2&\sim&\sum_i t_{3i}^2\vi h_i^{0r}\sim \varphi_3,\\
\phi_3&\sim&\sum_i[t_i(t_i+1)-t_{3i}^2]\vi h_i^{0r}\sim 
\varphi_4\equiv\phi_4. 
\eear
The quartic couplings can be read off the quartic potential for the 
Goldstone bosons
\be
\delta V = \frac{1}{24} \lambda_2 G_0^4 + \frac{1}{2}\lambda_3 G_0^2 G^+G^-
+\frac{1}{4}\lambda_4(G^+G^-)^2,
\ee
where
\be
G_0=\frac{g}{M_Z\cos\theta_W}\sum_i t_{3i} \vi h_i^{0i},
\ee
and
\be
G^+=\frac{g}{\sqrt{2}M_W}\sum_{ij}[(t^+_{ij}\vj)\Phi_i-\Phi_i^\dagger
(t^-_{ij}\vj)],
\ee
with $\Phi_i=(h_i^{0r}+ih_i^{0i})/\sqrt{2}$ as usual.

The final form of the bounds is
\bear
\label{bound2}
M_h^2&\leq&\frac{1}{3}\lambda_2 (v_o^2+v_e^2),\\
\label{bound3}
M_h^2&\leq&\lambda_3\rho\frac{1}{\sqrt{2}G_F}\equiv 
\lambda_3\rho v_W^2,\\ \label{bound4}
M_h^2&\leq&\frac{1}{2}\lambda_4 (v_o^2+v_e^2).
\eear
Here $\rho=M_W^2/(M_Z^2\cos\theta_W)$ and $v_W^2\geq 
v_o^2+v_e^2$.

{\bf Bound 5.}
The one we derived in Section 1. It has
\be
\label{field5}
\varphi_5\equiv\phi_5=\frac{1}{v_o}\sum_i{}^{'}\vi h_i^{0r},
\ee
with
\be
\delta V=\frac{1}{8}\lambda_5 \phi_5^4,
\ee
and reads
\be
\label{bound5}
M_h^2\leq\lambda_5 v_o^2.
\ee

It can be easily shown that for electroweak breaking driven only by 
doublets (so that $\rho=1$ automatically) all the bounds are exactly the 
same. In that case $v_e=0$ and $\phi_1=\phi_5$ (implying 
$\lambda_1=\lambda_5$) and the bounds (\ref{bound1}) and 
(\ref{bound5}) coincide trivially. To see that also the other three 
bounds are the same, notice that a suitable rotation in field space will 
permit us to write the doublet responsible of electroweak breaking as
\be
\Phi = \left(\begin{array}{c}
G^+\\
\frac{1}{\sqrt{2}}(v_o + \phi_1)+ \frac{i}{\sqrt{2}}G^0
\end{array}\right),
\ee
while the rest of doublets will play no role. Then the quartic coupling 
for $\Phi$ is written as
\be
\delta V=\frac{1}{2}\lambda(\Phi^\dagger\Phi)^2,
\ee
and expanding in components one gets\footnote{Actually this holds 
whenever the Higgs potential has a custodial $SU(2)_{L+R}$ symmetry (to 
ensure $\rho=1$).} \be 
\lambda=\frac{1}{3}\lambda_2=\lambda_3=\frac{1}{2}\lambda_4. \ee
Inserting these relations in the expressions (\ref{bound2}-\ref{bound4}) 
we recover always the result $\lambda v_o^2$.

In the most general case, with multiplets of different kinds 
contributing to the breaking, the bounds will be different and one has to 
choose the stronger. In addition, for particular models with extra 
symmetries, further bounds can be derived which may compete with the ones 
given here. 
\newline

{\bf 4. The decoupling limit}
\vspace{0.5cm}

From the mass bounds written in the form (\ref{intbound}) some extra 
useful information can be extracted. Consider one particular bound 
with $\phi_i=\varphi_i\equiv\phi$ so that (\ref{intbound}) reads \be
\label{decbound}
\sum_A M_A^2 |\langle \phi | A \rangle |^2 \leq \lambda v^2.
\ee
As noted in refs.~\cite{lanwel,wel} large masses can enter the sum 
(\ref{decbound}) and then
the corresponding eigenstates will have a small overlapping with $\phi$:
\be
|\langle \phi | A \rangle |^2 \leq \lambda \frac{v^2}{M_A^2}.
\ee
When all scalars but one are much heavier than the electroweak scale 
($M_H\gg v$),
eq. (\ref{decbound}) tells that the light state $|1\rangle$ is predominantly 
\footnote{The mass of this light Higgs is usually maximized in 
this limit. Note however that, even if $\langle\phi|M^2|\phi\rangle=\lambda 
v^2$, the mass will not saturate the bound 
$\lambda v^2$ necessarily.} $|\phi\rangle$: 
\be
|\langle\phi|1\rangle|^2=1-{\cal O}(v^2/M_H^2).
\ee
This in turn determines the properties of the light scalar in this 
decoupling limit. This issue has been studied in the literature 
concentrating mainly in the two-doublet model \cite{thdec}, where one 
finds that the light Higgs has standard couplings making extremely 
difficult to unravel the non minimal structure of the Higgs sector. 

Here we will concentrate on a different aspect of this problem. Suppose 
that $v_e$ is non-zero ({\em e.g.} some non-doublets contribute to 
electroweak breaking) so that the bounds {\em 1} and {\em 5} in the 
previous section are different. In such a case, what is the composition of 
the light Higgs in the decoupling limit? 
The first bound will give $|1\rangle=|\phi_1\rangle$ while the fifth 
will rather say $|1\rangle=|\phi_5\rangle$. 

The way out of this paradox 
is that the decoupling limit cannot be reached in that case, that is,
one cannot arrange the model so that all the scalars but one are heavy.
In other words, the combination of two different bounds, associated with two 
different field space directions, should provide a bound on the 
second-to-lightest scalar particle. This bound can be easily derived. Let 
us call $M_{1,2}^2$ the squared masses of the two light scalars 
($M_2^2>M_1^2$) and  $\alpha_{15}$ the angle ($0<\alpha_{15}\leq \pi/2$) 
between the two field directions $\phi_1$ and $\phi_5$. Next, $P|1\rangle$ 
is the (normalized) projection of the lightest scalar eigenstate onto the 
plane spanned by $\phi_1$ and $\phi_5$. Call $\beta$ the angle 
($-\pi/2\leq \beta\leq \pi/2$) between $P|1\rangle$ and $|\phi_1\rangle$.
Examining the quantities $\langle\phi_1|M^2-M_1^2|\phi_1\rangle$ and
$\langle\phi_5|M^2-M_1^2|\phi_5\rangle$ we get the inequalities:
\bear
(M_2^2-M_1^2)\corl 1 - |\langle \phi_1 | 1 
\rangle|^2\corr&\leq&\lambda_1v_1^2- M_1^2,\nonumber\vspace{0.2cm}\\
(M_2^2-M_1^2)\corl 1 - |\langle \phi_5 | 1 
\rangle|^2\corr&\leq&\lambda_5v_5^2- M_1^2.
\eear
Noting that 
\bear
|\langle\phi_1|1\rangle|^2&\leq&|\langle\phi_1|P|1\rangle|^2
=\cos^2\beta,\nonumber\vspace{0.2cm}\\
|\langle\phi_5|1\rangle|^2&\leq&|\langle\phi_5|P|1\rangle|^2
=\cos^2(\alpha_{15}-\beta),
\eear
it then follows
\bear
\label{boundnext}
M_2^2-M_1^2&\leq&\max_{|1\rangle}\llavl \min\corrl\frac{\lambda_1 
v_1^2 - M_1^2}{1-|\langle\phi_1|1\rangle|^2},\frac{\lambda_5 
v_5^2 - M_1^2}{1-|\langle\phi_5|1\rangle|^2}\corrr\llavr 
\nonumber\vspace{0.2cm}\\
&\leq& \max_{-\pi/2\leq\beta\leq\pi/2}\llavl \min\corrl\frac{\lambda_1 
v_1^2 - M_1^2}{\sin^2(\alpha_{15}-\beta)},\frac{\lambda_5 
v_5^2 - M_1^2}{\sin^2\beta}\corrr\llavr \vspace{0.2cm}\\
&=&\frac{1}{\sin^2\alpha_{15}}\llavl\lambda_1v_1^2+\lambda_5v_5^2-2M_1^2 
+ 2 
\corl(\lambda_1v_1^2-M_1^2)(\lambda_5v_5^2-M_1^2)\cos^2\alpha_{15}
\corr^{1/2}\llavr\nonumber.
\eear 
Or going even further
\bear
M_2^2-M_1^2&\leq&\frac{1}{\sin^2\alpha_{15}}
\llavl\corl\lambda_1v_1^2-M_1^2\corr^{1/2}
+\corl\lambda_5v_5^2-M_1^2\corr^{1/2} 
\llavr^2\sim\frac{\lambda v^2}{\sin^2\alpha}\nonumber.
\eear 
As is clear from this last expression the bound disappears for 
$\sin\alpha\rightarrow 0$ which corresponds to the situation of 
bounds with equal field directions (or, in the particular case we 
analyzed, to $v_e\rightarrow 0$). In practice, for 
$\sin^2\alpha\sim\lambda v^2/M_H^2$ the mass of the second-to-lightest Higgs 
can (in principle) be as heavy as $M_H$ and the decoupling limit with only 
one light Higgs can be realized.
Note that, in the derivation of (\ref{boundnext}) the presence of 
$\alpha_{15}$ in the denominator $\sin^2(\alpha_{15}-\beta)$, is crucial
to avoid the possibility of both denominators going to zero simultaneously
in which case the bound would be lost. 

It is clear that one can combine any two independent bounds in the same 
way as we have done and obtain different bounds on $M_2^2$. By 
independent bounds we mean bounds with linearly independent associated 
field directions\footnote{For example, bounds  2, 3, 4 are not linearly
independent. Moreover, it can be shown that, if two independent bounds 
$\langle\phi_a|M^2|\phi_a\rangle\leq \lambda_a v_a^2$ and
$\langle\phi_b|M^2|\phi_b\rangle\leq \lambda_b v_b^2$ exist, the 
off-diagonal matrix element satisfies 
$\langle\phi_a|M^2|\phi_b\rangle\leq \sqrt{\lambda_a\lambda_b} v_a v_b$.
For this reason, in this section we have restricted the discussion to 
diagonal bounds. Off-diagonal bounds, like bound~3 of the previous section
are spurious.}.
It should be clear then that from three independent 
bounds a limit on the mass of 
the third light scalar can be extracted. 
In general, having N bounds 
${\cal B}_a$, with associated field directions $\varphi_a=\phi_a$, the 
masses of M lighter scalars can be bounded, with $M=rank\{\phi_a\}\leq N$.
\newline

{\bf 5. Production Cross Sections} 
\vspace{0.5cm} 

Besides putting limits on the scalar masses it is important to know the 
composition of the light states in order to see if they can be produced 
at all in accelerators. The paradigmatic situation is exemplified by the 
next to minimal 
Supersymmetric Standard Model, NMSSM, which contains an extra chiral 
singlet in addition to the MSSM particle content. As is well known, a 
light scalar should be present in the spectrum of this model, but it can 
be singlet dominated and then hard to produce. In an 
interesting paper, ref.~\cite{kot}, it was shown that in the 
circumstance of 
the lightest Higgs boson being predominantly a singlet the second to 
lightest Higgs scalar will have an upper mass bound not far from the 
original 
bound on the lightest Higgs mass. And in case that also this second Higgs 
is singlet dominated, the third scalar will be subject to a similar bound.
This led to the conclusion that one of the three scalars will be produced 
in a future $e^+e^-$ linear collider (operating at $\sqrt{s}\sim 300\ 
GeV$) abundantly enough to guarantee detection. In this section we will 
show that such ladder of upper bounds can be generalized and applies to 
all models we are considering. The detectability at a Next Linear 
Collider can be studied in particular cases using these results and 
following the same procedure of Ref.~\cite{kot} (see also \cite{kingwhite}).

To derive the bounds we use the field $\phi_3$ of section 3. The 
superposition of a scalar eigenstate $H_i^0$ with $\phi_3$ is a 
measure of the strength of the gauge coupling of that eigenstate:
\be
\langle\phi_3|H_i^0\rangle\sim\frac{ZZH_i^0}{ZZH_{SM}}.
\ee 
Then consider the quantities $\langle\phi_3|M^2-m_N^2|\phi_3\rangle$, 
where $m_N^2$ is the squared mass of the $Nth$ scalar eigenstate.
For $N=1$ we reproduce the bound of section 3:
\be
m_1^2\leq\lambda_3v_3^2.
\ee
For N=2 is easy to get the inequality
\be
m_2^2\leq\frac{\lambda_3v_3^2-V_{11}^2m_1^2}{1-V_{11}^2},
\ee
where $V_{11}^2=|\langle H_1^0|\phi_3\rangle|^2$.
In general one obtains for the $Nth$ eigenstate 
\be
m_N^2\leq\frac{\lambda_3v_3^2-\Sigma_N^2m_1^2}{1-\Sigma_N^2},
\ee
with 
\be
\Sigma_N^2=\sum_{p=1}^{N-1}V_{1p}^2=\sum_{p=1}^{N-1}|\langle 
H_p^0|\phi_3\rangle|^2.
\ee
As explained in ref.~\cite{kot} the limit of small $\Sigma_N$ 
corresponds to the case of the first $N-1$ scalars being mostly decoupled 
from the $Z$ boson. In that case the bound on 
the $Nth$ scalar is stronger. 

From these mass bounds, and knowing the couplings to the Z boson one can 
put lower bounds on the production cross sections and study the 
capabilities of future $e^+e^-$ linear colliders for the discovery of one 
of the neutral scalars of a given model.  
\newline 

{\bf 7. Conclusions} 
\vspace{0.5cm} 

We have presented a novel and simple bound on the (tree-level) mass of the 
lightest 
neutral scalar in multi Higgs electroweak models. Whenever the Higgs 
sector is weakly coupled the bound presented implies the existence of 
a scalar state at or below the scale of symmetry breaking. This result
has applications in many different models of electroweak symmetry 
breaking and can be generalized to any model in which a continuous 
symmetry group is spontaneously broken. Here we have restricted our 
attention, as an example, to general low-energy supersymmetric models. 
We have reproduced some 
previous analytical bounds in a straightforward manner and clarified 
the limits  of applicability of the usual method to extract the bound in 
such models. 

Also, by comparing our new bound with previous general bounds existent in 
the literature, we have shown how to extract in some cases  extra 
information on the mass of heavier neutral scalar states. 
Some implications for the so-called decoupling limit in multi Higgs 
models have been obtained.

Finally we have generalized previous interesting results of Kamoshita et al. 
(obtained for the NMSSM, ref.~\cite{kot}) concerning the detectability of 
light neutral Higgses at the NLC.
\newline

{\bf Acknowledgements}
\vspace{0.5cm} 

We thank Howie Haber for discussions and very useful suggestions.
\newline

{\bf Appendix} 
\vspace{0.5cm}

We present here an example of supersymmetric model for which the 
usual technique to extract an "electroweak-controlled" bound on the 
mass of the lightest Higgs boson (see e.g. Ref.~\cite{kane}) would fail. 
The field content of the model is that of the MSSM supplemented by a 
chiral $SU(2)_L$ triplet ${\hat \Sigma}$, with hypercharge 1, and a 
4-plet ${\hat \Xi}$ with $Y=-1/2$, plus all extra fields necessary to 
cancel anomalies. The Yukawa couplings and VEVs of these extra fields 
will be assumed to be negligible so that we will ignore them in the 
following, its presence being unimportant for the effect we want to discuss. 

Let us write the fields $\Sigma_{ij}$ and $\Xi_{ijk}$ ($i,j,k=1,2$) in the
following matrix representation
\bear
\nonumber
\Sigma_{ij}&=&\left(
\begin{array}{cc}
\sigma_2^+/\sqrt{2}&-\sigma_1^{++}\\
\sigma_3^0&-\sigma_2^+/\sqrt{2}
\end{array}\right),
\eear
\bear
\Xi_{1ij}=\left(
\begin{array}{cc}
-\xi_2^0/\sqrt{3}&\xi^+_1\\
-\xi_3^-/\sqrt{3}&\xi_2^0/\sqrt{3}
\end{array}\right)\, &,& \Xi_{2ij}=\left(
\begin{array}{cc}
-\xi_3^-/\sqrt{3}&\xi^0_2/\sqrt{3}\\
-\xi_4^{--}&\xi_3^-/\sqrt{3}
\end{array}\right).
\eear
The superpotential contains a part
\be
\delta f=\epsilon_{ij}\parl\mu H_{1i} H_{2j} + \lambda H_{1i} \Sigma_{jk} 
H_{1k} + \sqrt{3} \zeta H_{1i} \Xi_{jkl} \Sigma_{lk}\parr
+3\gamma\Sigma_{ij}\Xi_{1jk}\Xi_{2ki},
\ee
and for the neutral components we get
\be
\label{neusup}
f=\mu h_1^0 h_2^0 + \lambda \sigma_3^0 h_1^0 h_1^0  + \zeta h_1^0 \sigma_3^0 
\xi_2^0 +\gamma\xi_2^0\xi_2^0\sigma_3^0
\ee
from which we can obtain the potential for $h_1^0, h_2^0, \sigma_3^0, 
\xi_2^0$. In general these fields will develop VEVs $v_1,v_2,s_3,x_2$ and, 
to satisfy the experimental constraints on $\Delta \rho$, one has to 
impose either that $s_3^2, x_2^2 \ll v_1^2+v_2^2$ or $s_3^2 \simeq 3 x_2^2$.
 
In any case, it is straightforward to obtain the mass matrix for the real 
fields $h_1^{0r},h_2^{0r},\sigma_3^{0r},\xi_2^{0r}$ in the vacuum 
$(v_1,v_2,s_3,x_2)$. Looking only to the $h_1^{0r},h_2^{0r}$ submatrix 
one gets (neglecting gauge contributions for simplicity):
\bear
M^2_{11}&=&-(B + 2\lambda s_3) \mu\tan\beta -
 \zeta A s_3 \frac{x_2}{v_1} \nonumber\\
&+& 4\lambda^2 v_1^2 + 3\lambda\zeta v_1 x_2 - \zeta\gamma\frac{x_2^3}{v_1}
-2(\lambda+\gamma)\zeta s_3^2 
\frac{x_2}{v_1},\nonumber\\
M^2_{12}&=& (B + 2\lambda s_3)\mu,\\
M^2_{22}&=&-( B + 2\lambda s_3) \mu\cot\beta -
\zeta \mu s_3 \frac{x_2}{v_2}\nonumber,
\eear 
where $B,A$ are soft masses. We see that this matrix is not of the general 
form 
obtained in \cite{kane} although the model satisfies all the conditions
required in that paper. The expressions given above show clearly that the 
bound one can obtain from this $2\times 2$ matrix is not controlled by the
electroweak scale.

The situation resembles that encountered (see \cite{pandita}) 
in the Supersymmetric Singlet Majoron Model \cite{gmpr} and it has a similar 
solution \cite{espinosa}. Let us redefine the fields\footnote{Note that 
$\xi_2^{0r}$ and $h_2^{0r}$ belong to different $SU(2)_L$ 
representations so that (\ref{newf}) is not a mere 
redefinition of the "correct" doublet fields.}
\bear
\label{newf}
\phi_1&=&h_1^{0r},\nonumber\\
\phi_2&=&\frac{1}{v'_2}(v_2 h_2^{0r} + x_2 \xi_2^{0r}),
\eear
where $v'_2{}^2=v_2^2+x_2^2$. The mass submatrix for $\phi_1,\phi_2$ has 
now the well-behaved form 
\begin{eqnarray}
M'{}^2&=&\left|\left|
\begin{array}{cc}
-m'_3{}^2\tan\beta'+ \Delta_{11}&
m'_3{}^2 + \Delta_{12}\\
&\\
m'_3{}^2 + \Delta_{12}&
-m'_3{}^2\cot\beta'
+ \Delta_{22}
 \\
\end{array}\right|\right|,
\end{eqnarray}
with 
\bear
m'_3{}^2=( B + 2\lambda s_3) \frac{v_2}{v'_2} +
\zeta A s_3 \frac{x_2}{v'_2} &,& \tan\beta'=\frac{v'_2}{v_1},
\eear
and
\bear
\Delta_{11}&=&4\lambda^2 v_1^2+\frac{x_2}{v_1}\corl 3\lambda\zeta v_1^2  
-\zeta\gamma x_2^2 - 2(\lambda+\gamma) \zeta s_3^2 \corr ,\nonumber\\
\Delta_{12}&=&\frac{x_2}{v'_2}\corl
2(\zeta^2+2\gamma\lambda)v_1x_2+\zeta\lambda (3v_1^2+2s_3^2)+
\zeta\gamma (3x_2^2+2s_3^2)\corr,\\
\Delta_{22}&=&\frac{x_2}{v'_2{}^2}\corl4\gamma^2x_2^3+
\zeta\gamma(3x_2^2-2s_3^2)v_1-\zeta\lambda(2s_3^2+v_1^2)v_1
\corr
\nonumber. \eear
From this matrix the Yukawa-part of the mass bound can be obtained as
\be
\label{yukcota}
\delta m_h^2= \Delta_{11}\cos^2\beta' + \Delta_{22}\sin^2\beta' 
+\Delta_{12}\sin 2\beta'= 4 (\lambda c_1^2 + \zeta c_1 c_x + \gamma c_x^2
)^2 v'{}^2 ,
\ee
with $v'{}^2=v_1^2+v_2^2+x_2^2$, $c_i=v_i/v'$, $c_x=x_2/v'$. Of course,
this is the result one can obtain using the simple prescription given in 
the text. Defining $\phi^0=c_1 h_1^{0}+c_2 h_2^{0} + c_x \xi_2^{0}$ and 
recasting the superpotential (\ref{neusup}) in the form
\be
f=(\lambda c_1^2 + \zeta c_1 c_x +\gamma c_x^2) \sigma_3^0 \phi^0 \phi^0 +...
\ee
gives directly the bound (\ref{yukcota}).


\begin{thebibliography}{99}
%
\bibitem{lanwel}
P. Langacker and H.A. Weldon, \PRL 52 84 1377.
%
\bibitem{wel}
H.A. Weldon, \PLB 146 84 59.
%
\bibitem{bucce} F. Buccella, G.B. Gelmini, A. Masiero and M. Roncadelli, 
\NPB 231 84 493.
%
\bibitem{eq} J.R. Espinosa and M. Quir\'os, \PLB 279 92 92;{} \PLB 302 93 51.
%
\bibitem{kane} G. Kane, C. Kolda and J. D. Wells, \PRL 70 93 2686.
%
\bibitem{espinosa} J.R. Espinosa, \PLB 353 95 243.
%
\bibitem{pandita} P.N. Pandita, \MPLA 10 95 1533.
%
\bibitem{gmpr} G.F. Giudice, A. Masiero, M. 
Pietroni and A. Riotto, \NPB 396 93 243; \\ 
 M. Shiraishi, I. Umemura and K. Yamamoto, \PLB 313 93 89.
%
\bibitem{thdec} H. Georgi and D. Nanopoulos, \PLB 82 79 95;\\
H.E. Haber and Y. Nir, \NPB 335 90 363;  \\
H.E. Haber, [hep-ph/9501320] and [hep-ph/9505240].
%
\bibitem{kot}
J. Kamoshita, Y. Okada and M. Tanaka, \PLB 328 94 67.
%
\bibitem{kingwhite}
S.F. King and P.L. White, \PRD 53 96 4049.
%
\end{thebibliography}
\end{document}